\documentclass[aps,prb,twocolumn,superscriptaddress,floatfix,amsmath,amssymb,showpacs]{revtex4}

\usepackage[dvips]{graphicx}
\usepackage[dvipdfm]{hyperref}

\begin{document}

\title{Antiferromagnetic Order in MnO Spherical Nanoparticles}

\author{C. H. Wang}
\affiliation{Oak Ridge National Laboratory, Oak Ridge, Tennessee 37831, USA}
\author{S. N. Baker}
\altaffiliation[Current Address: ]{Department of Chemical Engineering, University of Missouri, Columbia, MO 65211, USA}
\affiliation{Oak Ridge National Laboratory, Oak Ridge, Tennessee 37831, USA}
\author{M. D. Lumsden}
\author{S. E. Nagler}
\author{W. T. Heller}
\affiliation{Oak Ridge National Laboratory, Oak Ridge, Tennessee 37831, USA}
\author{G. A. Baker}
\altaffiliation[Current Address: ]{Department of Chemistry, University of Missouri, Columbia, MO 65211, USA}
\affiliation{Oak Ridge National Laboratory, Oak Ridge, Tennessee 37831, USA}

\author{P. D. Deen}
\altaffiliation[Current Address: ]{European Spallation Source ESS AB, Stora Algatan 4, SE-221 00 Lund, Sweden}
\affiliation{Institut-Laue Langevin, 6 Rue Jules Horowitz, 38042 Grenoble, Cedex 9, France}

\author{L. M. D. Cranswick}
\affiliation{Canadian Neutron Beam Centre, National Research Council of Canada, Chalk River Laboratories, Chalk River, Ontario, K0J 1J0, Canada}

\author{Y. Su}
\affiliation{J\"{u}lich Centre for Neutron Science, Forschungszentrum J\"{u}lich, Outstation at FRM II, Lichtenbergstrasse 1, 85747 Garching, Germany}

\author{A. D. Christianson}
\affiliation{Oak Ridge National Laboratory, Oak Ridge, Tennessee 37831, USA}

\date{\today}

\begin{abstract}

We have performed unpolarized and polarized neutron diffraction experiments on
monodisperse 8 nm and 13 nm antiferromagnetic MnO nanoparticles. For the 8 nm sample,
the antiferromagnetic transition temperature $T_N$ (114 K) is suppressed
compared to the bulk material (119 K) while for the 13 nm sample $T_N$ (120 K)
is comparable to the bulk. The neutron diffraction data of the nanoparticles is well described using the bulk MnO magnetic
structure but with a substantially reduced average magnetic moment
of 4.2$\pm$0.3 $\mu_B$/Mn for the 8 nm sample and 3.9$\pm$0.2 $\mu_B$/Mn for the 13 nm sample.
An analysis of the polarized neutron data on both samples shows that in an individual MnO nanoparticle
about 80$\%$ of Mn ions order. These results can be explained by a structure in which the monodisperse
nanoparticles studied here have a core that behaves similar to the bulk with a surface layer which
does not contribute significantly to the magnetic order.

\end{abstract}


\maketitle

\section{Introduction}

The fundamental magnetic behavior at the nanoscale has attracted considerable
attention due to potential technological applications such as magnetic data storage\cite{data}
and energy storage lithium ion batteries.\cite{batteries}
Because of the finite size of nanoscale magnetic materials, the large surface-to-core ratio
becomes significant and, in some cases, explains the fascinating nanoscale behavior. The number
of nearest neighbors for an atom at the surface is significantly
reduced compared to the bulk with the consequence that the magnetic exchange is lessened.
Consequently, the magnetization and the transition temperature to magnetic order may be substantially
reduced when compared to the bulk. Theoretical calculations have predicted this
behavior.\cite{Merikoski,Hendriksen,Lopez} Experimentally, this behavior has been
observed in both 3$d$ metals\cite{Heer1} and oxide samples such as MnO\cite{Sako1,Sako2} and
NiO.\cite{Klausen} In contrast, an increase has been observed in the magnetic ordering temperature
and magnetic moment in several nanoscale materials.\cite{Billas,Golosovsky,Golosovsky1,Golosovsky2,Ortega,Taylor}
For example, in some nanoscale 3$d$ metals, the moment is enhanced as a result of band narrowing at the surface
resulting from the reduced coordination number. This behavior has been
observed experimentally in iron, cobalt and nickel clusters\cite{Billas} and has been supported
by theoretical calculations.\cite{Pastor} On the other hand, in ionic oxide compounds, due to the
relatively localized electronic distribution the moment is less affected by the surface and surface disorder
results in a reduced average moment.\cite{Kodama} A reduced moment
and enhanced transition temperature have been observed in MnO
particles\cite{Golosovsky,Golosovsky2,Golosovsky1,Ortega,Taylor} and in the spinel ferrites
Ni(Mn)Fe$_2$O$_4$.\cite{MFe2O4} In either case, the origin of the moment and ordering temperature
enhancement or reduction is unclear and, consequently, further study of magnetic nanoparticles is
important.

The classic antiferromagnetic (AFM) oxide compound MnO is a good candidate for studying finite
size effects due to its relatively simple structure and well-studied bulk properties.
Bulk MnO shows an AFM transition at $T_N$ $\approx$
118 K-120 K\cite{Shull,Morosin,Bonfante} that occurs concomitantly with a rhombohedral structural
distortion from the high temperature NaCl structure.\cite{Tombes}  Previous studies have shown that
the antiferromagnetic order and structural transition of MnO survive into the nanoscale regime.\cite{Golosovsky}
The nature of the effect and its root cause remain unresolved because in some studies an enhanced $T_N$ has been
observed for confinement particles\cite{Golosovsky,Golosovsky2,Feygenson} and core-shell
particles\cite{Ortega,Golosovsky1} while a suppressed $T_N$ is reported in so-called ultrafine
particles.\cite{Sako1,Sako2} In another example of MnO nanoparticles, only short range AFM order is
reported.\cite{Chatterji} The degree to which this difference in behavior is related
to the magnetic domain size or to the surface preparation remains unclear. Hence, studies that
compare multiple sizes of nanoscale MnO with similar surface preparations can shed light on the
intrinsic magnetic behavior that results from size confinement.

In this paper we report neutron diffraction measurements on two different sizes (8 nm and 13 nm)
of monodisperse spherical  MnO nanoparticles.
The 8 (13) nm sample is denoted as sample A (B) in the following discussion.
Our results show that in sample A the AFM transition temperature $T_N$ is suppressed while in sample B,
$T_N$ is almost the same as bulk MnO. The Mn magnetic moment is similar in both samples and is about 80$\%$
of the bulk value of 4.89 $\mu_B$/Mn.\cite{Bonfante} From polarized neutron diffraction
data we estimate that the surface to total ratio is about 20 \% which appears to explain the reduction
in the moment.

\section{Experimental Details}

Monodisperse MnO nanoparticles were synthesized through a modification of non-injection
synthetic schemes.\cite{Sun} An important aspect of the synthesis process is the attachment of capping ligands
which provides size control, minimizes interparticle interactions, and passivates the surface.  To minimize
difficulties associated with the large incoherent neutron scattering cross-section of hydrogen, deuterated
capping ligands were used.  Approximately 10 batches for each sample size were combined to produce samples large enough for neutron scattering experiments($\sim$ 0.5 g each). Two distinct sizes were studied, an 8 nm sample (Sample A) and a 13 nm (sample B).  These particle sizes are determined from transmission electron microscopy (TEM) of $\sim$50-100 particles per sample.  As will be discussed further below, a limitation of TEM is the low number of particles sampled, however despite this limitation the TEM can provide some information concerning the polydispersity of the samples studied here. The TEM results on sample A give a particle size of 7.9 $\pm$ 1.6 nm (ref. 25) and for sample B give a particle size of 13 $\pm$ 2 nm (see insets Fig. 3(a) and (b)).  The shape of nanoparticle samples is an important consideration.  Aside from the difficulty in forming anisotropic particles of MnO using our methods, there are two primary lines of evidence for the production of spherical particles. The first comes from TEM images. If other shapes are formed, projection in two dimensions will reveal other geometries. For example, cubic nanoparticles will occasionally (for suitable orientation) reveal hexagonal shapes based on the 2D projection. Additionally, if plates are formed, these are generally transparent in TEM imaging and this was not the case here. Furthermore, if growth occurs along a particular crystallographic face, this will be revealed in X-ray and neutron diffraction patterns. Indeed, preferred growth directions will be associated with different peak widths for different reflections. In this case, no such divergence was found, consistent with the formation of isotropic (\textit{i.e.}, spherical) particles, in strong corroboration of our TEM results.

X-ray diffraction with Cu $K\alpha$ radiation ($\lambda$=1.5406 {\AA}),
and polarized and unpolarized neutron diffraction experiments
were performed. The neutron diffraction experiments were performed on several instruments: the HB1A triple-axis
instrument with $\lambda$ = 2.364 {\AA} at the High Flux Isotope Reactor (HFIR), Oak Ridge National
Laboratory; the C2 diffractometer at the Canadian
Neutron Beam Centre (NBC) in Chalk River, Canada, with neutron wavelength of $\lambda$ = 1.3306 {\AA}
and 2.3721 {\AA}; the general purpose neutron polarization analysis
spectrometer, D7, with $\lambda$ = 3.073 {\AA} at the Institut Laue-Langevin in Grenoble, France;
and the DNS polarized diffuse scattering instrument with $\lambda$ = 4.74 {\AA}
operated by the J\"{u}lich Centre for Neutron Science (JCNS) at the Forschungsneutronenquelle Heinz Maier-Leibnitz
(FRM II), TU M\"{u}nchen, Germany.

To carefully extract the particle size as well as the magnetic domain size from a diffraction
pattern, it is important to properly account for instrumental resolution.
A standard run on a mixture of NIST Si 640c and annealed Yttria was performed to determine the instrumental
resolution of C2.  Similarly, a Y$_3$Fe$_5$O$_{12}$ standard was used to determine the instrumental resolution of the D7. For HB1A the
instrumental resolution was obtained by comparing with a Si standard and with a sample of bulk MnO.
The particle size and magnetic domain size were obtained from analysis of the (1 1 1) and (1/2 1/2 1/2)(pseudo cubic notation)
peak width using the standard Scherrer formula\cite{Scherrer} and refinement of the diffraction pattern
using Fullprof.\cite{Fullfrof} In particular, two methods were used to extract the particle size and magnetic domain size from the neutron diffraction data.  In the first method,
the instrumental resolution obtained as described above was convolved with the fitted Gaussian peak width.  In the second method, the instrument resolution peak shape parameters were fixed and a peak broadening parameter, to account for finite size effects, was included in the Fullprof refinements of the data.  All particle sizes determined from the neutron
diffraction data are reported with the instrumental resolution taken into account.

\section{Results and analysis}



\subsection{Sample A (8 nm)}

Fig 1(a) shows the x-ray diffraction pattern of sample A. The diffuse background is likely a consequence
of the capping ligand. Analysis of the widths of the
(111), (002) and (022) reflections using the Scherrer formula\cite{Scherrer} yields particle
sizes of 9.5 $\pm$ 0.2 nm, 9.7 $\pm$ 0.1 nm and 9.2 $\pm$ 0.1 nm, similar to the values
determined from transmission electron microscopy (TEM) of 7.9 $\pm$ 1.6 nm.\cite{Sun}  Note that in contrast to the TEM data, the error bars on the particle size determined from the x-ray data (as well as the neutron data discussed below) reflect a statistical error of the mean particle size and do not provide information concerning the polydispersity of the samples.
In Fig. 1(c) we present neutron diffraction data collected on the triple-axis instrument HB1A with
$\lambda$ = 2.3639 {\AA} at 20 K. As shown in Fig. 1(b), the peak width of the nanoparticle sample is obviously
broadened when compared to the bulk material.  The particle size determined from the nuclear peak(accounting
for instrumental resolution) is  9.9 $\pm$ 0.5 nm in good agreement with the x-ray values.  Here we note that
the amount of material used differs significantly for the different techniques.  TEM measurements sample
the fewest particles ($\sim$ 50-100 particles) followed by x-ray diffraction ($\sim$ 50 mg) and, finally, neutron diffraction ($\sim$ 0.5 g).
Given this, we use the neutron derived values of the particle size as being most representative of the samples
measured.  In a similar fashion, the effective magnetic domain size can be determined from the width of
the magnetic Bragg peaks.  In sample A
at 20 K this analysis yields a magnetic domain size of 9.2 $\pm$ 0.4 nm.  Within the experimental resolution
of this measurement we are unable to determine a shift in lattice constant compared to the bulk, however
the x-ray data (higher resolution) on samples from this batch indicate a somewhat smaller lattice
constant (4.439(1) {\AA}).

\begin{figure}[t]
\centering
\includegraphics[width=0.4\textwidth]{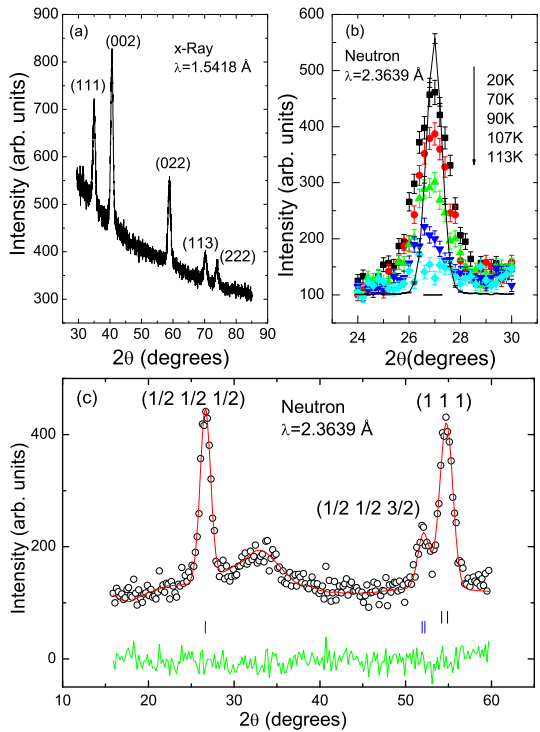}
\caption{\label{fig1} (Color online) (a) x-ray diffraction pattern of MnO nanoparticle sample A.
(b) The temperature dependence  of the (1/2 1/2 1/2)   magnetic Bragg peak for sample A.
The solid line is the scaled data for the bulk material and the horizontal
bar indicates the full width at half maximum (FWHM) of the instrumental resolution.
(c) Neutron diffraction data for sample A at 20 K. The data
were collected on HB1A. The broad hump between 30 to 40 degrees is likely the contribution from
the capping ligand and/or residual solvent. This hump is also observed in the coherent + isotope incoherent
scattering channel for sample B at $Q \sim$ 1.5 {\AA} (see Fig. 3(c)), indicating that this scattering is of nonmagnetic origin.
The solid line is the result of the Rietveld refinement. The short vertical lines indicate the nuclear Bragg peak (upper) and magnetic Bragg peak (lower).}
\vspace*{-3.5mm}
\end{figure}

In Fig. 2, we present the temperature dependence of the magnetic moment for both bulk MnO and sample A. No significant thermal hysteresis was observed in either the bulk or nanoparticle material indicating that the samples were in thermal equilibrium.  The
moment was obtained by scaling the integrated intensities of magnetic and structural Bragg reflections,
assuming a moment of 4.89 $\mu_B$/Mn\cite{Bonfante} for bulk MnO.
This results in a low temperature magnetic moment for sample A of
4.07 $\pm$ 0.11 $\mu_B$/Mn at 20 K. Note that this analysis assumes the magnetic structure of the nanoparticle and the bulk is the same aside from the magnitude of the ordered
magnetic moment-justification of this assumption is given below.
From Fig. 2, we can clearly see that not only is the
magnetic moment suppressed in the nanoparticle sample relative to the bulk, but that the AFM transition
temperature $T_N$ decreases from 118.7 K in bulk to 113.6 K  in sample A. A similar suppression
of $T_N$ has been reported in ultrafine MnO particles with sizes in the range of $\sim$3.7 to $\sim$ 5.4 nm.\cite{Sako1,Sako2}

\begin{figure}[t]
\centering
\includegraphics[width=0.9\columnwidth]{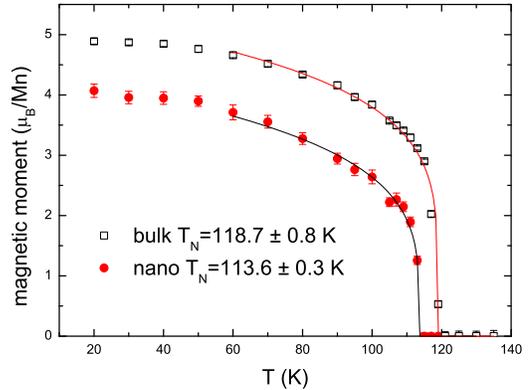}
\caption{\label{fig2} (Color online) The temperature dependent magnetic moment of the MnO bulk
and nanoparticles of sample A. The solid lines are guides for the eye. }
\vspace*{-3.5mm}
\end{figure}

To provide further insight into the above results, models of both the crystal and the magnetic structure were fit to the neutron diffraction data of sample A and a sample of bulk MnO at 20 K, through Rietveld refinement using Fullprof to determine the magnetic moment, particle size, and domain size.
A model including the rhombohedral structural distortion (R $\overline{3}$ m)\cite{Roth} was
used for the refinement and the bulk diffraction pattern was used to specify the instrument resolution
function for the sample A refinement. Fitting the bulk neutron scattering data yields a magnetic moment of 4.8 $\pm$ 0.3 $\mu_B$/Mn, consistent with the previously reported value.\cite{Bonfante} The resulting refinement for sample A is indicated by the solid line in
Fig. 1 (c). The results yield a magnetic moment of 4.2 $\pm$ 0.3 $\mu_B$/Mn, a particle size of 10.3 $\pm$ 0.5 nm
and a magnetic domain size of 9.1 $\pm$ 0.7 nm, consistent with the analysis above using the Scherrer formula.  Here we note that the magnetic domain size is almost as large as the nuclear particle size ($\sim$1.5 magnetic domains per nanoparticle), thus the dominant magnetic behavior appears to come from a single magnetic domain per nanoparticle, this point will be discussed further below.


\subsection{Sample B (13 nm)}

\begin{figure}[t]
\centering
\includegraphics[width=0.95\columnwidth]{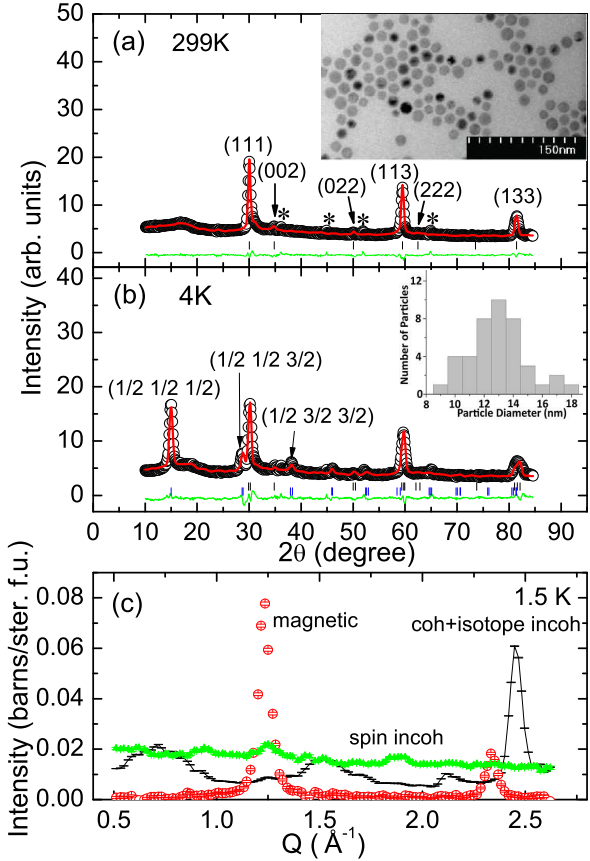}
\caption{\label{fig3} (Color online) The observed (open circles) and calculated (solid lines) neutron diffraction
data for nanoparticle sample B at room temperature (a) and 4 K (b). The weak peaks indicated
by the star in (a) are due to a MnO$_2$ impurity. The data were collected on C2.
The short vertical lines indicates the nuclear (upper) and magnetic (lower) Bragg Peak position.
(c) The separated polarized neutron diffraction data. The intensity is presented in units of barns per steradian per formula unit. The data were collected on D7. The solid line is the
sum of the coherent and isotope incoherent, the solid triangle is nuclear spin incoherent scattering and the open circle
is the magnetic scattering. The inset in (a) is the TEM image.  The inset in (b) is the particle size distribution.}
\vspace*{-3.5mm}
\end{figure}

In Fig. 3(a) room temperature diffraction data are presented for sample B. Analyzing the (1 1 1) nuclear reflection
yields a particle size of 13.1 $\pm$ 0.7 nm. Fig. 3(b) shows the
diffraction pattern at 4 K. A similar analysis of the width of the magnetic peak (1/2 1/2 1/2)
gives a magnetic domain size of 10.0 $\pm$ 0.2 nm. Polarized neutron diffraction data allows for the unambiguous
separation of the magnetic from the nonmagnetic scattering. In Fig. 3(c) we present
polarized neutron diffraction data separated into various components of the neutron scattering cross-section.
Here the single peak analysis yields a magnetic domain size of 9.7 $\pm$ 0.4 nm and a
 particle size of 12.6 $\pm$ 0.8 nm.  Thus in this case, there appears to be the possibility of 2 magnetic domains per nanoparticle.

To thoroughly analyze the diffraction data, Fullprof was used for the refinement of the entire diffraction
pattern. The solid lines in Fig. 3(a) and (b) are the results of the refinement. As in the bulk
material, at a temperature above the AFM transition the diffraction pattern can be well-described
by a NaCl structure (F m $\overline{3}$ m) while below the transition temperature, a structural
model including a rhombohedral distortion (R $\overline{3}$ m) best describes the data.

\begin{figure}[t]
\centering
\includegraphics[width=0.85\columnwidth]{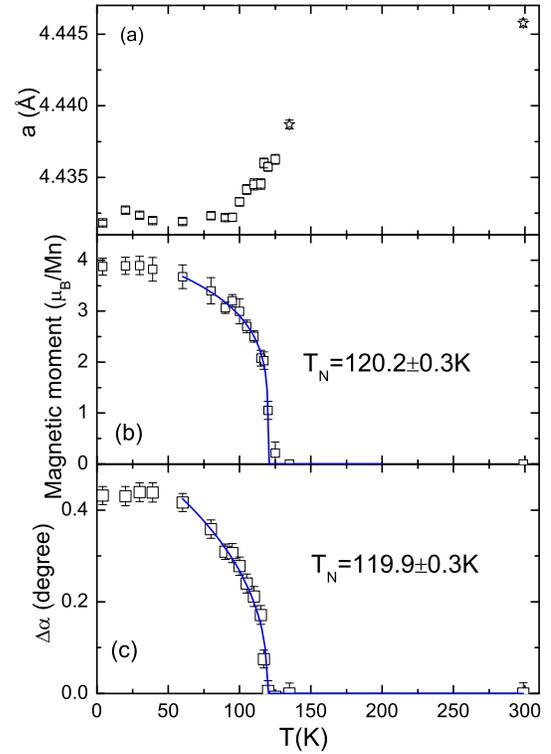}
\caption{\label{fig4} (Color online) Temperature dependence of the lattice parameter (a)
magnetic moment (b) and the structure distortion angle, $\Delta \alpha$ (c) for sample B. The results are
obtained through refinement of models of the crystal and magnetic structure  to the neutron scattering data from C2 as described in the text.
The solid lines in (b) and (c) are guides for the eye. }
\vspace*{-3.5mm}
\end{figure}

In Fig. 4 we present the results of structural refinements as a function of temperature for sample B.
The temperature dependent lattice parameter is shown in Fig. 4(a); the stars are the results for the high temperature NaCl
phase. The temperature dependence of the  magnetic moment is shown in Fig. 4(b). This allows us to extract an
AFM transition temperature of 120 $\pm$ 0.3 K, comparable, or perhaps slightly larger than that of the bulk material.
At 4 K the magnetic moment saturates at a value of 3.9 $\pm$ 0.2 $\mu_B$/Mn, which is similar to the previous reported value for
MnO nanoparticles embedded in porous glass.\cite{Golosovsky} Fig. 4(c) shows the temperature dependence of the rhombohedral
distortion angle, $\Delta \alpha$. $\Delta \alpha$ increases with decreasing temperature saturating at a value of
0.44 $\pm$ 0.01$^o$. This value is slightly smaller than both the nanoparticles embedded in porous glass\cite{Golosovsky}
and the bulk\cite{Morosin} where $\Delta \alpha$ $\sim$ 0.6$^o$. However, a smaller value of about 0.43$^o$
has also been reported in bulk.\cite{Roth} The temperature dependence of the distortion angle indicates a structural transition
temperature of 120 $\pm$ 0.3 K, which is the
same as the $T_N$ obtained from the evolution of the magnetic order parameter.

\begin{figure}[t]
\centering
\includegraphics[width=0.4\textwidth]{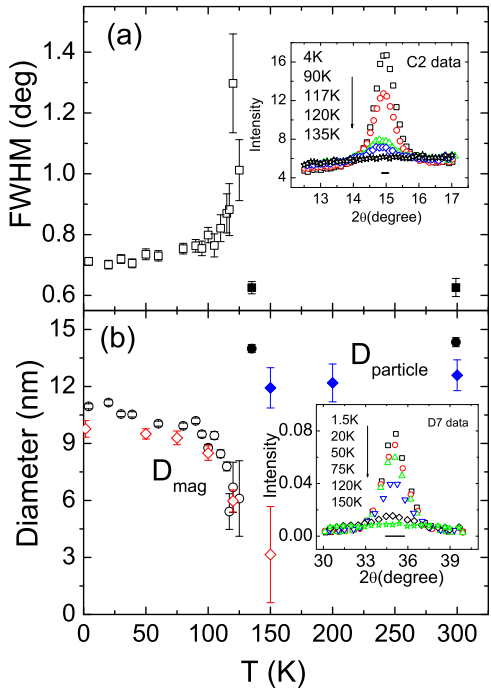}
\caption{\label{fig5} (Color online) (a) (1/2 1/2 1/2) magnetic peak width (open symbol) and the (1 1 1) nuclear peak width (solid symbol)
for sample B. Inset is the (1/2 1/2 1/2) peak at different temperatures. The data were collected on the C2 instrument.
(b) The temperature dependent magnetic domain size (open symbols) and particle size (solid symbols) for
sample B. Circles are unpolarized results from C2 and diamonds are polarized results from D7. Inset is the
evolution of the (1/2 1/2 1/2) magnetic peak with temperature from the D7 polarized neutron scattering data. Horizontal bars in the insets represent the
instrumental resolution. }
\vspace*{-3.5mm}
\end{figure}

The temperature evolution of the (1/2 1/2 1/2)   magnetic Bragg peak is displayed in
the insets of Fig. 5(a) and Fig. 5(b), respectively. In Fig. 5(a) the temperature dependence of the full
width at half maximum (FWHM) of the (1/2 1/2 1/2) peak is shown for sample B. Above $T_N$, there is no obvious
signal at (1/2 1/2 1/2); we include the (1 1 1) nuclear peak width for comparison (solid symbol). The magnetic
peak width shows a drastic change at $T_N$ and becomes progressively narrower below $T_N$ indicating that the magnetic
domains emerge and grow progressively with decreasing temperature. The temperature dependence of the refined average
domain diameter is shown in Fig. 5(b).  For comparison, above $T_N$, the
average particle size is plotted (solid symbols). A similar domain size is determined from the polarized neutron scattering data.
The results obtained from both C2 (unpolarized) and D7 (polarized) are plotted together in Fig. 5(b) for comparison.

\section{discussion}

The summary of the particle and magnetic parameters are listed in Table I.
Both samples exhibit a suppressed magnetic moment compared to bulk MnO.
This behavior is also observed in confinement geometry and core-shell MnO
particles\cite{Golosovsky,Ortega,Golosovsky2} and is believed to be the result of
disordered spins on the surface of the nanoparticles.\cite{Golosovsky,Kodama}
However, whether the actual moment
for the Mn site is the same as in the bulk has not been directly determined yet. To test this,
polarized neutron scattering data which is sensitive to both the ordered and
disordered moment is required. Previous polarized neutron scattering
data on average 7 nm confinement geometry nanoparticles indicated that only 40$\%$ of the Mn atoms
were ordered.\cite{Feygenson} However, the average magnetic moment is not reported. In the polarized
data presented here on sample B at 1.5 K, from the peak-to-diffuse scattering ratio we can estimate that about
80-85$\%$ of the Mn atoms in a nanoparticle are magnetically ordered. Assuming the actual magnetic moment of Mn atoms in the nanoparticle is identical to the bulk value and scaling by the percent of the particle that actually participates in magnetically ordered state, the expected average moment in this sample would be 3.9-4.1 $\mu_B$/Mn. This is consistent with the values of the magnetic moment derived from the neutron scattering data reported in Table I. In sample A, the average magnetic moment is about 85$\%$ of the bulk material. The polarized diffraction results collected using DNS (not shown) show that about 75-80$\%$ Mn atoms order at 3 K, again, implying that the ordered moment of Mn site is similar to the bulk. One possible explanation for this observation is that the  nanoparticles studied here contain a core which acts like bulk MnO and a shell with random spins.

\begin{table}[htp]
\caption{\label{tab:table} Summary of the particle and magnetic parameters obtained through x-ray, TEM
and neutron scattering analysis. D$_{particle}$ and D$_{mag}$ are the diameter of particles and magnetic domain size respectively.
Error bars in the x-ray and neutron diffraction results represent statistical standard deviation $\pm \sigma$ of the mean particle size while the
TEM results represent the standard deviation of the size distribution.}
\begin{ruledtabular}
\begin{tabular}[b]{ccc}
 samples    &  A  &  B \\
\hline
D$_{particle}$ (nm) &  10.3 $\pm$ 0.5 (HB1A) & 14.0 $\pm$ 0.2 (C2)\\
                   & 9.5 $\pm$ 0.2 (x-ray)& 12.6 $\pm$ 0.8 (D7)  \\
                   & 7.9 $\pm$ 1.6 (TEM) &   13.0 $\pm$ 2.0 (TEM)\\
D$_{mag}$ (nm) & 9.1 $\pm$ 0.7 (HB1A) &  10.9 $\pm$ 0.1 (C2)  \\
   &            & 9.7 $\pm$ 0.4 (D7) \\
$T_N$ (K)  &  113.6 $\pm$ 0.2 (HB1A) & 120.2 $\pm$ 0.3 (C2)  \\
moment ($\mu_B$/Mn) & 4.2 $\pm$ 0.3 (HB1A) &  3.9 $\pm$ 0.2 (C2) \\
        &             & 3.9 $\pm$ 0.2 (D7) \\
\end{tabular}
\vspace{-2mm}
\end{ruledtabular}
\end{table}

The nature of the surface spins has been discussed frequently in the literature.  In particular, magnetic susceptibility measurements have been interpreted in terms of weak ferromagnetism and related superparamagnetic like behavior of uncompensated surface spins.\cite{Ghosh, Morales,Lee,Seo}  However, the size dependence of the superparamagnetic behavior appears to be inversely related to particle size in contrast to the direct relationship expected for a superparamagnetic particle.  Morales, \textit{et al.}\cite{Morales} have proposed that the Mn residing on the surface is subject to a noncubic crystalline electric field and this provides an additional source of anisotropy effecting the superparamagnetic behavior. In this lower symmetry crystal field environment, a low-spin configuration is energetically favorable resulting in a moment of $\sim 1 \mu_B$ on the surface Mn atoms.  Unfortunately, the neutron scattering data here can not contribute significantly to the discussion concerning the nature of the surface spins.  The diffuse magnetic scattering in the polarized neutron scattering data (Fig. 3(c)) shows a weak magnetic diffuse component that suggests the presence of disordered spins, but we stress though that the measurement itself provides no information whether or not these spins are on the surface.  Moreover, if there were significant ferromagnetic moments we would expect depolarization of the neutron beam.  No obvious depolarization of the neutron beam was observed, but it is difficult make a quantitative estimate on a limit of the a ferromagnetic moment size without further systematic measurements.

%

A comparison of T$_N$ for bulk MnO to that of sample A (8 nm), shows that T$_N$ is suppressed by about 4$\%$
compared to the bulk value.  On the other hand, sample B (13 nm), exhibits a $T_N$ comparable to the
bulk. Suppression of $T_N$ is also observed in ultrafine MnO particles \cite{Sako1} as well as in other nanoscale
AFM compounds such as NiO and CoO thin films,\cite{NiO,CoO} and NiO disc shaped nanoparticles.\cite{Klausen}
A mean field approach with finite size effects for the magnetic transition has been used for the explanation of
the suppression of $T_N$. In this theory, the transition temperature is suppressed when the sample
size/thickness decreases to small value. For MnO a prism model of ultrafine particles, the calculations
predict $T_N$ will be suppressed at small size due to the decrease of the average coupling constant which
is induced by the decrease of coordination number on the surface.\cite{Sako1} Similar arguments are likely to
apply to the nanoparticles studied here and thus $T_N$ suppression is likely due to decreases in the
coordination number and the average coupling constant. In sample B, $T_N$ remains the same as bulk
material suggesting that size confinement effects in MnO nanoparticles appear at only relatively small sizes.  In apparent contradiction to the preceding arguments, an enhanced $T_N$ is reported in confined geometry nanoparticles.\cite{Golosovsky,Ortega,Golosovsky2,Feygenson}
For these nanoparticle systems the worm-like particle morphology\cite{Golosovsky,Golosovsky2,Feygenson} and the surface interface of glass\cite{Golosovsky,Golosovsky2,Feygenson} or another magnetic species\cite{Ortega} is the likely cause of the differences in behavior with those studied here. As mentioned above, the coordination number on the surface affects the transition
temperature and by adjusting the coordination number T$_N$ can be increased or decreased.\cite{Wesselinowa}

The effect of size confinement on the order of the magnetic/structural phase transition is another
interesting issue in nanoscale MnO.  It has been suggested in the confined geometry MnO nanoparticles that the AFM transition is
a continuous phase transition.\cite{Golosovsky,Golosovsky2} In these samples the critical exponent
$\beta$ describing the ordered magnetic moment,
M(T) $\sim$ (1-$T/T_N$)$^{\beta}$,  ranges from 0.3 up to almost the mean field value of
0.5. These values are in reasonable agreement with Monte Carlo simulations for the finite size Ising
model and Heisenberg model where $\beta$ is 0.3258 and 0.3616, respectively.\cite{Landau}
In the reduced temperature range of 1-$T$/$T_N$= 0.02-0.5, fitting our temperature dependent
moment to a power law of (1-$T$/$T_N$)$^\beta$ yields a much smaller $\beta$ of
0.24$\pm$0.02 for sample A and 0.22$\pm$0.02 for sample B.  This could be interpreted as lower dimensional
critical behavior but is more likely the result of a discontinuous phase transition.

In conclusion, we have studied two different sizes of monodisperse MnO nanoparticles using
unpolarized and polarized neutron diffraction. Both the magnetic and crystal structure are similar to the bulk,
but with substantially reduced average magnetic moment.  Moreover, the results show that in the
8 nm sample the antiferromagnetic transition temperature $T_N$ is suppressed
while in the 13 nm sample, $T_N$ is comparable to bulk. The suppression of $T_N$ is attributed
to the exchange coupling reduction induced by the finite size effect. The observations
presented here are consistent with the core of the MnO nanoparticles behaving much like the bulk with
a disordered surface layer.

\section{Acknowledgments}
We acknowledge useful discussions with J. Musfeldt and I. Swainson.
Research Work at ORNL was sponsored by the Laboratory Directed Research
and Development Program of ORNL, and was supported by the
Scientific User Facilities Division Office of Basic Energy Sciences,
DOE. A portion of this research was conducted at the Center for Nanophase Materials
Sciences, which is sponsored at ORNL by the Office of Basic Energy Sciences, U.S. DOE.

\end{document}